\documentclass[prl,twocolumn,amsmath,amssymb,superscriptaddress]{revtex4-1}
\usepackage{graphicx}
\usepackage{bm}
\usepackage{mathptmx}
\usepackage[english]{babel}
\usepackage{indentfirst}
\usepackage{color}
\usepackage{listings}
\usepackage{multibib}
\usepackage{hyperref}
\hypersetup{
     colorlinks   = true,
     citecolor    = blue,
     linkcolor    = blue
           }

\begin{document}
\bibliographystyle{unsrt}

\title{A majority gate with chiral magnetic solitons}

\author{Konstantinos Koumpouras}
\affiliation{Department of Physics and Astronomy, Uppsala University, Box 516, SE-751 20 Uppsala, Sweden}

\author{Dmitry Yudin}
\affiliation{ITMO University, Saint Petersburg 197101, Russia}

\author{Christoph Adelmann}
\affiliation{Imec, Kapeldreef 75, B-3001 Leuven, Belgium}

\author{Anders Bergman}
\affiliation{Department of Physics and Astronomy, Uppsala University, Box 516, SE-751 20 Uppsala, Sweden}
\affiliation{Maison de la Simulation, CEA, CNRS, Univ. Paris-Sud, UVSQ, Universit{\'e} Paris-Saclay, F-91191 Gif-sur-Yvette, France}
\affiliation{INAC-MEM, CEA, F-38000 Grenoble, France}

\author{Olle Eriksson}
\affiliation{Department of Physics and Astronomy, Uppsala University, Box 516, SE-751 20 Uppsala, Sweden}
\affiliation{School of Science and Technology, \"Orebro University, SE-70182 \"Orebro, Sweden}

\author{Manuel Pereiro}
\email[Corresponding author. ]{manuel.pereiro@physics.uu.se}
\affiliation{Department of Physics and Astronomy, Uppsala University, Box 516, SE-751 20 Uppsala, Sweden}

\maketitle

{\bf 
In magnetic materials nontrivial spin textures may emerge owing to the competition among different types of magnetic interactions. Among such spin textures, chiral magnetic solitons represent topologically protected spin configurations with particle-like properties. Based on atomistic spin dynamics simulations, we demonstrate that these chiral magnetic solitons are ideal to use for logical operations, and we demonstrate the functionality of a three-input majority gate, in which the input states can be controlled by applying an external electromagnetic field or spin-polarized currents. One of the main advantages of the proposed device is that the input and output signals are encoded in the chirality of solitons, that may be moved, allowing to perform logical operations using only minute electric currents. As an example we illustrate how the three input majority gate can be used to perform logical relations, such as Boolean AND and OR.}


For several decades, progress in micro- and nanoelectronics has been driven by the miniaturization of the underlying fundamental building blocks, complementary metal--oxide--semiconductor (CMOS) transistors and interconnects. In recent years, this process has become increasingly cumbersome and both material \cite{chau2007integrated} as well as device \cite{ferain2011multigate,palacios2012applied} innovations had to be introduced. Although this process is poised to continue for at least another decade \cite{kuhn2012considerations}, novel technology beyond CMOS devices and circuits are being actively researched to add complementary functionality to future nanoelectronic circuits and to eventually replace CMOS \cite{bernstein2010device,nikonov2013overview} when its fundamental intrinsic limitations will be reached \cite{markov2014limits}.

Two major attributes of desired devices that go beyond CMOS technology are low power dissipation and nonvolatility. In this context, spintronic devices that are based on spin instead of charge degrees of freedom are of particular interest since devices based on collective (ferro-)magnetic states and their excitations may allow to simultaneously reduce power and provide nonvolatility. In addition, spintronic devices are especially suitable for building majority gates \cite{radu2015spintronic}. A majority gate is a logical gate that returns true if and only if more than half of its inputs are true, otherwise the output is false (see Table \ref{truth_table}). Circuit design based on majority gates in combination with inverters has recently received much interest since it has been shown that it can lead to smaller, faster, and more energy efficient circuits compared to traditional approaches for AND and OR logic gates \cite{amaru2015new, amaru2015new2}. Hence, majority gates can be considered to be key devices in novel nanoelectronic circuit architectures with improved area and power scaling behaviour, something we explore here.

To date, spintronic majority gates based on nanomagnetic logic \cite{imre2006majority}, domain wall movement \cite{nikonov2011proposal}, as well as spin wave interference \cite{khitun2008spin, klingler2014design} have been proposed. The recent rapid progress in the development of fabrication and characterization methods of magnetic nanostructures \cite{sellmyer2006advanced} have brought spintronic majority gates close to their realization. However, much conceptional work, for example on interconnect schemes \cite{dutta2015non}, is still needed before such devices can be integrated in large scale circuits. Very recently, magnetic skyrmions and solitons \cite{kang,nagaosa2013} have attracted much attention because of their topological stability \cite{fert} and the ability to manipulate them by spin polarised currents \cite{yu}. Such magnetic solitons belong to the class of non-collinear magnetic structures, where the origin of the non-collinearity lies in the competition between Heisenberg exchange and the Dzyaloshinskii-Moriya interaction (DMI) \cite{dzyaloshinsky1958thermodynamic, moriya1960anisotropic}. Their inherent topological stability makes these particle-like objects stable and prevents them from collapsing in a short time span, while their propagation velocity makes them highly suitable for fast nanoelectronic logical gates \cite{pereiro}. Moreover, magnetic skyrmions can be manipulated with ultralow electric current densities ($10^{-12}$ A/nm$^2$) that are more than $10^5$ times smaller than those used to move other magnetic textures, such as domain walls \cite{yu-kanazawa}. This is because skyrmions can couple efficiently to the current and are not strongly affected by disorder in atomic magnetic moments. 

\section{Description of the majority gate device}
\begin{figure}[ht]
	\centering\includegraphics[scale=0.18, angle=270]{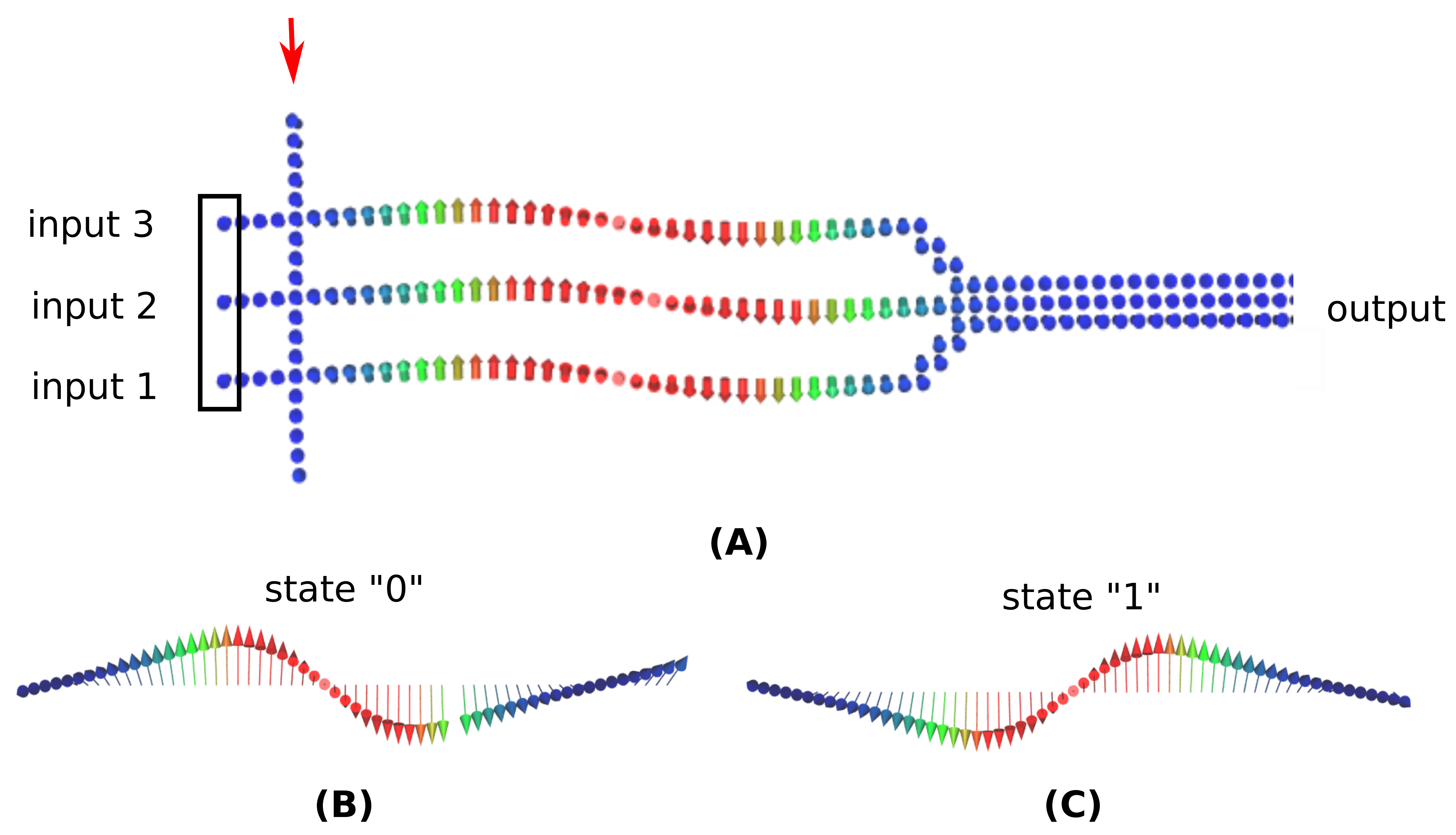}
	\caption{\label{device} \textbf{(A)} Snapshot of the solitonic majority gate device where on the left there are three input arms and on the right one single output arm. Perpendicular to the input arms, an arm used to generate the topological solitons is shown, in which the direction of a spin-polarized current is indicated by a red arrow. In the text this is referred to as the soliton generating arm. One soliton is stabilised in each branch with input states: "0" for input 1, "0" for input 2 and "0" for input 3. 
\textbf{(B, C)} Zoomed-in snapshots of the two different logic states "0" (left image) and "1" (right image). The direction of the applied external magnetic field is perpendicular to the plane of the figure and points towards the reader. The blue arrows represent atomic magnetic moments, that are aligned along the $\hat{\mathbf{z}}$ axis parallel to the direction of the external field, whereas the red ones are atomic moments that form a collinear configuration opposite to the direction of the external magnetic field. A left winding soliton is assigned a logical state  "0" (as shown in the left image-(\textbf{B)}) and a right winding soliton a logical state  "1" (as shown in the right image-(\textbf{C})) .}
\end{figure}

In this paper, we propose the usage of chiral magnetic solitons as information carriers in a majority gate. The information is coded in the chirality of the soliton (see Figs.~\ref{device}b-c). More concretely, the helicity $\gamma = - \pi/2$ corresponds to the logic state ``0", whereas $\gamma = \pi/2$ corresponds to logic ``1". Recent experimental and theoretical findings have identified the chiral helimagnet Cr$_{1/3}$NbS$_2$ \cite{moriya1982evidence} as a plausible candidate to support a quasi-one-dimensional soliton lattice that is stabilised by an external magnetic field \cite{togawa2012chiral}. The main focus of this paper is to extend the physical understanding developed for these quasi-one-dimensional solitons and to demonstrate how one may use them in the novel concept of solitonic majority gates. 

In contrast to widely used conventional ferromagnetic materials, a delicate interplay between Heisenberg exchange, DMI, and anisotropic interaction can favor the formation of non-collinear ordering in chiral helimagnets. Interestingly, the presence of an external magnetic field perpendicular to the helical axis stabilizes the soliton lattice \cite{togawa2012chiral}. Such particle-like solitons, which are the ground states of non-linear field models, can be manipulated, controlled, and used to transfer information. Moreover, they can be driven efficiently by comparatively low currents, as we will show below.

\begin{table}[h]
	\caption{\label{truth_table} Table of the logic operations for the chiral, solitonic majority gate. When the gate called input 1 is "0", the device performs an AND- operation and when it is "1" an OR- operation.}
	\begin{center}
		\begin{tabular}{|c|c|c||c||c|c|c|c|}
			\hline \hline
			Input 1 & Input 2 & Input 3 & Output &Input 1 & Input 2 & Input 3 & Output \\
			\hline \hline
			\multicolumn{4}{|c||}{AND-} & \multicolumn{4}{|c|}{OR-}\\
			\hline
			0 & 0 & 0 & 0 & 1 & 1 & 0 & 1\\
			0 & 0 & 1 & 0 & 1 & 0 & 1 & 1\\
			0 & 1 & 0 & 0 & 1 & 0 & 0 &0\\
			0 & 1 & 1 & 1 & 1 & 1 & 1 & 1\\
			\hline 
			\hline
		\end{tabular}
	\end{center}
\end{table}

We choose to describe the spin system of chiral magnets by a classical atomistic spin-Hamiltonian, that contains Heisenberg exchange, uniaxial magnetic anisotropy, DMI and a Zeeman term (for details, see Methods). 
We have investigated the magnetization dynamics via an atomistic Landau-Lifshitz-Gilbert equation, using the UppASD package~\cite{skubic2008method}. In this formalism it is possible to study the influence of spin-polarised electrical currents. Details of our simulations can be found in the Methods section.


We now explore a device structure for magnonics that relies on chiral magnetic solitons. The device can be described as a chiral, solitonic majority gate and we outline here a practical protocol for its implementation. We propose a setup that consists of three input arms which allow for the formation and propagation of solitons, the central region where the solitons interact to perform computation and the output region (schematically shown in Fig.~\ref{device}a in which each arrow represents the direction of the magnetic moment on each atomic site).
We assume that Heisenberg exchange, $J$, as well as DMI, $D$, act among spins in three inputs, whereas the
output arm
lacks DMI and is characterized by a smaller value of exchange coupling $J^\prime$ (in the Supplementary Material we consider a case with DMI in the output arm, as a way to detect the chirality of the output signal, this is described in more detail below). 
 The Heisenberg exchange is assumed to be of nearest neighbor type, with a strength of $J=1$ mRy. We note here, that the functionality of the majority gate is not dependent on this assumption. 

\section{Majority gate functionality}
The solitons are created by applying a spin polarised current through the soliton generating arm which is perpendicular to the three input arms as shown by the red arrow in Fig.~\ref{device}a. When a current flows through this soliton generating arm with current density $j=3.9 \times 10^{12}$ A/m$^2$, after a very short time ($t= 25$ ps) solitons are generated in each input arm (see Supplementary Movie 1). An alternative way of generating solitons is by applying a local magnetic field, which locally reverses the magnetisation direction. When the local field is removed, the system relaxes and the solitons are stabilised (see Supplementary Movie 2). 


\begin{figure}[ht]
	\centering\includegraphics[scale=0.2, angle=0]{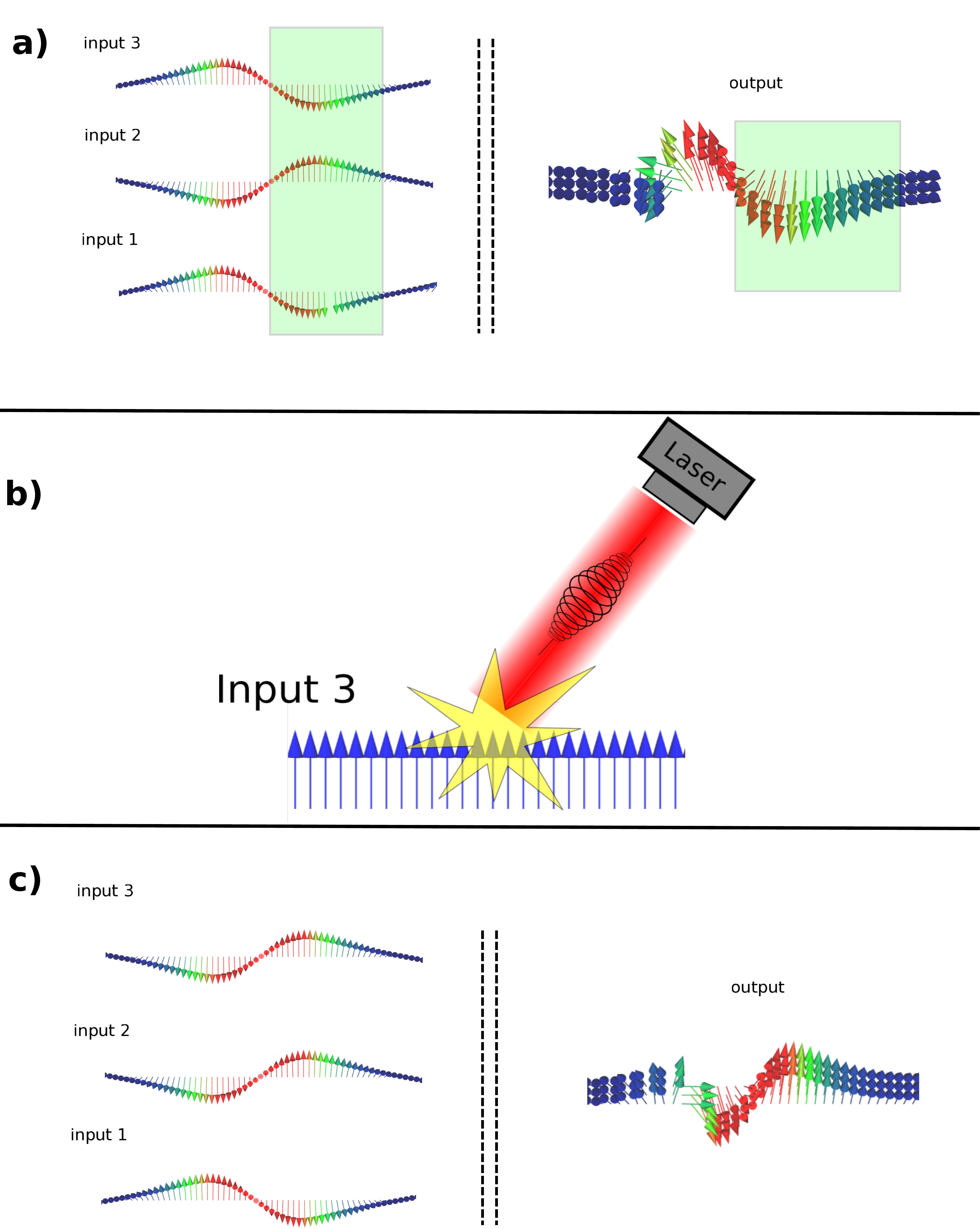}
	\caption{\label{input_output} Zoomed-in snapshots from ASD simulations which show the inputs, output and the procedure of reversing the chirality. \textbf{a)} Majority gate operation where the three inputs are "0", "1" and "0", and the output is "0". 
\textbf{b)} Zoomed-in figure of input arm 3, where an external circularly polarised electromagnetic field is applied in order to reverse the DM vector and create a soliton with opposite chirality. \textbf{c)} Generation of solitons after the electromagnetic field and the three inputs now are ``1", ``1" and ``0". The output in this case is ``1". }
\end{figure}

Having been generated in input arms 1,2 and 3, the solitons can be made to move under the influence of a very weak spin polarised current, as is shown schematically in Fig.~\ref{input_output}. We argued in the beginning of this report that the most important property of solitons for practical implementation of the majority gate is their chirality and particularly the region of the soliton which is shown from the shaded green boxes in Fig.~\ref{input_output}. In the output arm, where three solitons of the input arms meet, interaction among them takes place. Due to the absence of DMI in the output arm, the magnetic ground state is in accordance with Heisenberg exchange interaction $J^\prime$, which in this case means a collinear ferromagnet. However, a chiral magnetic excitation may nevertheless form in the output arm. Many simulations with different input files based on UppASD were carried out for the device shown in Fig.1, and it clearly manifest that chiral excitations can be stabilized in the output arm. Once the chiral solitons of the input arms are allowed to interact in the output arm, the result is always a chiral soliton that reflected the chirality of the majority of the three input arms. 
\section{Results}
Results of such simulations are shown e.g. in Fig.~\ref{input_output}a, where the three input lines carry the logical information ''0'', ''1'', ''0'' and the output carries the information ''0''. In  Fig.~\ref{input_output}c the three input lines have the logical information ''1'', ''1'', ''0'' and the output signal is a ''1''.
Based on our simulations, the truth table of the majority gate was established, as shown in Table I. 
Before continuing our discussion, we note as a detail, that the soliton of the output arm is smaller in size as compared to those in the input arms. This is natural, because the size of the soliton, defined as the distance between the reversed core of the soliton and the opposite ferromagnetic background outside of the soliton, depends in the input arm on the $D/J$ ratio while the edge of the merged soliton of the output arm depends on the $K/J^\prime$ ratio.

Note from Table I that if input arm 1 has input "0", the device works in Boolean AND mode, i.e. the output is a "1" only if input arm 2 and 3 both have "1" as input. Otherwise the output arm has a "0". This then defines the AND operation. On the other hand, OR operation takes place when the input arm 1 has input "1" and the output is "0" only if the other two input arms are "0". The control of the device allowed for by the input arm 1, enables a dynamical device that may be changed from AND to OR functionality. There are several ways to accomplish this, here we discuss one of them. 
The information needed to establish a truth table, as given in Table \ref{truth_table}, is encoded in the chirality of the soliton in each input arm, which is determined by the DMI exclusively. The sign and strength of the DMI can be tuned by applying an external  electromagnetic field (schematically shown in Fig.~\ref{input_output}b and discussed in more detail in the Supplementary Material).
This provides an ability to choose the input state (``0" or ``1") in each arm of Fig.~\ref{device} and this then allows for the majority gate to operate as a dynamical gate. 


In order to realize an efficient device, it is crucial to easily detect the chirality of the soliton of the output arm of Fig.~\ref{device}. A way to achieve this is described in the Supplementary Material (see also Supplementary Movies 3 and 4). Briefly described, this relies on extending the output arm with two additional branches that have finite, equal strength DM interaction but with opposite sign. Left chiral solitons are then stable only in the output arm that has DMI that supports this chirality, while right chiral solitons exist only on the other output arm. As the simulations described in the Supplementary Material show, this selects with extremely high reliability, output signals with a specific chirality and therefore provides exact knowledge of the logical state of the output arm. On a more detailed level, we note that our results rely to come extent on the value of the value of $J^{\prime}$, which should not be bigger than a threshold value in order to make the device fully functional and reliable (for more details see Supplementary Material, where this is analyzed in detail).


While the research activity in spintronics has historically been focused on magnetically ordered materials, non-collinear magnetic excitations, such as chiral solitons, have very recently started to achieve considerable attention. It can be expected that one outcome of these efforts is the emergence of nanostructures and devices based on more exotic magnetic states, that will enable logic operations, such as inverters or fan out gates, and memory elements. In the present article we demonstrate one such development, as realized by the three-input majority gate.
The device analyzed here is discussed to have a dynamic capability, enabling AND and OR functionality, depending on the DMI of input arm 1. 
Generally we visualise DMI as indirect antisymmetric coupling between localized magnetic moments induced by conduction electrons \cite{imamura}. These electrons are Rashba spin-orbit coupled owing to the lack of inversion symmetry, while the coupling among localized magnetic moments and conduction electrons is realized via Kondo-type exchange coupling. It was recently shown that when such a system is exposed by external electromagnetic wave the corresponding coupling constant is dramatically reduced \cite{Hettler,Ng,Lopez,Goldin,Kaminski,Nordlander,Shahbazyan}. Therefore, a rather standard laser setup can be used to tune the value of DMI and possibly also the sign of the DM in any of the input arms \cite{sato}, enabling a device that dynamically can be changed from one operation to the other. Although this is certainly feasible experimentally, it is by no means the only way to make the device in Fig.1 operate.
Concluding, we hope our research will lead to a mutual inspiration between theory and experiment in the field of data processing and storage as well as nanoscale magnetism that rely on chiral states.

\section{Methods}
We consider a one-dimensional spin system, which is described by the spin Hamiltonian with the DMI, uniaxial anisotropy and Zeeman term,

\begin{equation}
\begin{split}
{\cal{H}}=-\frac{1}{2}\sum\limits_{i\neq j}J_{ij}\mathbf{m}_i\cdot\mathbf{m}_j-\frac{1}{2}\sum\limits_{i\neq j}\mathbf{D}_{ij}\cdot\left(\mathbf{m}_i\times\mathbf{m}_j\right) \\
+ K \sum_i (\mathbf{m}_i \cdot \mathbf{\hat{e}}_K)^2 -\mathbf{B}\cdot\sum\limits_i\mathbf{m}_i
\end{split}
\end{equation}
where $i$ and $j$ are atomic indices, $\mathbf{m}_i$ stands for the classical atomic moment, $J_{ij}$ is the strength of the Heisenberg exchange interaction, $\mathbf{D}_{ij}$ is the Dzyaloshinskii vector, while $K$ and $\mathbf{\hat{e}}_K$ denote the uniaxial anisotropy constant and the direction of the easy axis respectively, and the last term $\mathbf{B}$ is the external applied magnetic field. In this paper we work out magnetisation dynamics based on model parameters but the parameters have been chosen in a range accessible for realistic materials (see main text).

This investigation of solitons dynamics introduced in the system is performed in terms of atomistic spin dynamics \cite{antropov1996spin}, as implemented in the UppASD package \cite{skubic2008method}. In order to study the motion of the solitons in one-dimensional structure in the presence of spin polarised current, we have performed atomistic spin dynamics simulations based on the Landau-Lifshitz-Gilbert \cite{Gilbert} equation with additional terms to describe the spin transfer torque effect \cite{slonczewski}:

\begin{eqnarray}\nonumber
\begin{split}
\frac{\partial\mathbf{m}_i}{\partial t}=-\frac{\gamma}{1+\alpha^2}\mathbf{m}_i\times\left(\mathbf{B}_i^\mathrm{eff}+\frac{\alpha}{m_i}\left(\mathbf{m}_i\times\mathbf{B}_i^\mathrm{eff}\right)\right) \\ 
+\frac{1+\alpha\beta}{1+\alpha^2}u_x\mathbf{m}_i\times\left(\mathbf{m}_i\times\frac{\partial\mathbf{m}_i}{\partial x}\right) \\ -\frac{\alpha-\beta}{1+\alpha^2}u_x\mathbf{m}_i\times\frac{\partial\mathbf{m}_i}{\partial x}
\end{split}
\end{eqnarray}

\noindent where $\alpha$ is the Gilbert damping coefficient, $\gamma$ the gyromagnetic ratio, $\mathbf{m}_i$ the magnetic moment, $\mathbf{B}_i^\mathrm{eff}$ the effective field, \noindent $\beta$ the non-adiabatic parameter and $u_x$ the velocity term. The effective field is given by

\begin{equation}
\mathbf{B}_i^\mathrm{eff}=\mathbf{B}_i+\mathbf{b}_i(T)
\end{equation}

\noindent where $\mathbf{B}_i=-\partial H/\partial\mathbf{m}_i$ and takes the interactions in the system and $\mathbf{b}_i$ accounts for temperature fluctuations via a random Gaussian shaped field.

The velocity term is in units of velocity (m/s or similar) and is related to the magnitude and the direction of the current $\mathbf{j}$, is proportional to the applied current and is equal to

\begin{equation}
u_x=\frac{g\mu_BPj}{2eM_s}
\end{equation}

\noindent where $j$ is the current density, $P$ the polarisation and $M_s$ the saturation magnetisation.

\section{Acknowledgements}
We acknowledge financial support from the KAW foundation
(projects 2013.0020 and 2012.0031), Swedish
Research Council (VR) and eSSENCE. AB acknowledges support from the CEA Enhanced Eurotalents program. The calculations were performed at the computational
facilities of NSC (Link\"oping University, Sweden) under a project administrated
by SNIC/SNAC. D.Y. acknowledges the support from RFBR project 16-32-60040. CA’s work was supported by imec’s Industrial Affiliation Program on Beyond CMOS devices.

\section{Author contributions}
K. K., O. E., C. A and M. P. conceived the project. D.Y and M. P. performed  the  theoretical  model  and  theoretical  analysis.    K.K.  performed  the  atomistic  spin  dynamics  calculations.   All the authors  analysed the data.  K.K., D.Y. and M.P. co-wrote the manuscript. All  authors  discussed  the  data  and  commented  on  the manuscript.

\section{Additional information}
Supplementary information is available in the online version of the paper. Reprints and permissions information is available online at www.nature.com/reprints. Correspondence and requests for materials should be addressed to K.K. and M.P.
\vspace{2cm}

$^*$Corresponding author. manuel.pereiro@physics.uu.se

\end{document}


\title{Supplementary Material: \\ 
A majority gate with chiral magnetic solitons}

\author{Konstantinos Koumpouras} 
\affiliation{Department of Physics and Astronomy, Uppsala University, Box 516, SE-751 20 Uppsala, Sweden}

\author{Dmitry Yudin}
\affiliation{ITMO University, Saint Petersburg 197101, Russia}

\author{Christoph Adelmann}
\affiliation{IMEC, Kapeldreef 75, B-3001 Leuven, Belgium}

\author{Anders Bergman}
\affiliation{Department of Physics and Astronomy, Uppsala University, Box 516, SE-751 20 Uppsala, Sweden}
\affiliation{Maison de la Simulation, CEA, CNRS, Univ. Paris-Sud, UVSQ, Universit{\'e} Paris-Saclay, F-91191 Gif-sur-Yvette, France}
\affiliation{INAC-MEM, CEA, F-38000 Grenoble, France}

\author{Olle Eriksson}
\affiliation{Department of Physics and Astronomy, Uppsala University, Box 516, SE-751 20 Uppsala, Sweden}
\affiliation{School of Science and Technology, \"Orebro University, SE-70182 \"Orebro, Sweden}

\author{Manuel Pereiro}
\email{manuel.pereiro@physics.uu.se}
\affiliation{Department of Physics and Astronomy, Uppsala University, Box 516, SE-751 20 Uppsala, Sweden}

\maketitle

\onecolumngrid
\setcounter{page}{1}
\setcounter{table}{0}
\setcounter{section}{0}
\setcounter{figure}{0}
\setcounter{equation}{0}
\renewcommand{\thepage}{\Roman{page}}
\renewcommand{\thesection}{S\arabic{section}}
\renewcommand{\thetable}{S\arabic{table}}
\renewcommand{\thefigure}{S\arabic{figure}}
\renewcommand{\theequation}{S\arabic{equation}}

\section{Tuning Dzyaloshinskii-Moriya coupling strength and chirality with an external field}

In the main text we argued that an external electromagnetic field, e.g. from a laser, could be used to tune the value of Dzyaloshinskii-Moriya interaction (DMI) strength. In this section starting from a model Hamiltonian we show that this is indeed the case. It is well known that DMI stems from the lack of inversion symmetry in the system and microscopically originates from spin-orbit interaction (SOI). For simplicity, but without loss of generality, in what follows we consider the Hamiltonian of a one-dimensional electron gas (note, that each input or output arm of the majority gate represents a quasi-one-dimensional system) in the presence of the Rashba SOI,

\begin{equation}\label{hamiltonian}
H=-\frac{\hbar^2}{2m}\nabla^2+\alpha\left(-i\hbar\nabla\times\hat{\mathbf{z}}\right)\cdot\bm{\sigma},
\end{equation}

\noindent where $m$ is the effective electron mass, $\alpha$ stands for the SOI strength, while $\bm{\sigma}=\left(\sigma_x,\sigma_y,\sigma_z\right)$ denotes the vector of Pauli matrices. Suppose that two localized magnetic moments are embedded in an electron gas, the mutual influence of these two spins on conduction electrons can be captured by an $sd-$interaction model,

\begin{equation}\label{sd}
H_\mathrm{int}=V\sum\limits_{i=1,2}\delta\left(\mathbf{r-R}_i\right)\mathbf{S}_i\cdot\bm{\sigma},
\end{equation}

\noindent here, $V$ corresponds to the strength of $sd-$interaction, $\mathbf{R}_{1,2}$ denote the positions of two localized spins $\mathbf{S}_{1,2}$. For a quasi-one-dimensional electron system working out the second-order correction with respect to $V$ results in an antisymmetric term in the Hamiltonian,

\begin{equation}\label{dm}
H_\mathrm{DM}=\mathbf{D}\cdot\left(\mathbf{S}_1\times\mathbf{S}_2\right),
\end{equation}

\noindent that can be identified with DMI. The Dzyaloshinskii vector, $\mathbf{D}$,  turns out to be determined by the relative distance between the spins $R=\vert\mathbf{R}_1-\mathbf{R}_2\vert$ with the absolute value $D=\vert\mathbf{D}\vert$ \cite{Imamura},

\begin{equation}\label{dzyaloshinskii}
D(R)=\frac{2mV^2}{\pi\hbar^2}\left[\mathrm{Si}(2qR)-\frac{\pi}{2}\right]\sin(2m\alpha R/\hbar^2),
\end{equation}

\noindent where $q=\sqrt{2mE_F+m^2\alpha^2}/\hbar$ is defined by the Fermi energy, $E_F$, and $\mathrm{Si}(z)=\int_0^z dt\sin t/t$ is the sine integral. Thus, Eq. (\ref{dzyaloshinskii}) allows one to estimate indirect DMI between a pair of spins induced by conduction electrons.

To complete our task we briefly recall that the strength of the $sd-$interaction can be evaluated from the microscopic Anderson impurity model (AIM) \cite{Hewson}. The AIM serves as one of the paradigmatic models in studies of strongly-correlated electron systems. Originally introduced to describe magnetic impurities in metal hosts it allows to explain the formation of localized magnetic moments. In the strong coupling regime the $sd-$Hamiltonian (\ref{hamiltonian}) can be derived from AIM by applying a Schriffer-Wollf transformation \cite{Hewson}. At the same time it was recently reported that when the system is exposed by an external monochromatic field the coupling constant $V$ becomes renormalized \cite{Kaminski}. Interestingly, for a high-frequency off-resonant external pumping (when the frequency of a laser sets the dominant energy scale in the system) this renormalization is given by

\begin{equation}\label{coupling}
\tilde{V}=VJ_0^2\left(\frac{eV_\omega}{\hbar\omega}\right).
\end{equation}

\noindent Here, $eV_\omega$ determines the amplitude of ac modulation, $\omega$ is the frequency of the field, and $J_0(z)$ denotes the zeroth order Bessel function of the first kind. Thus, the effect of exposing the system to a high-frequency laser field results in a renormalization of the corresponding coupling (\ref{coupling}). This value, when plugged into (\ref{dzyaloshinskii}), leads to a decrease of the DMI strength with the field (note that $J_0(z)\leq1$).

The chirality of DMI can be artificially generated and  tuned by using a circularly polarised laser pulse, as already proposed in Ref.~\cite{sato} by applying Floquet theory to a time-periodic spin Hamiltonian. Here the periodicity in time is generated by the time-dependence of the electromagnetic field. The laser-driven DMI is given by:
\begin{equation}\label{laser}
	H_{laser}=\frac{\alpha\beta}{2\omega}{\bf R}_{12} \cdot \left({\bf S}_1\times {\bf S}_2\right)\cos\delta
\end{equation}
with $\alpha=g_{me} F_0$, $\beta=g\mu_B F_0 c^{-1}$, ${\bf R}_{12}$ the vector connecting spins ${\bf S}_1$ and ${\bf S}_2$, and $\delta$ the helicity of the laser ($\delta=0$ and $\delta=\pi$ for right-circularly and left-circularly polarised laser, respectively). The speed of light is represented by c while $g_{me}$ and $F_0$ are the magnetoelectric coupling constant and intensity of the field, respectively. The frequency of the electromagnetic field is indicated by $\omega$. From Eq.~(\ref{laser}), it is clear that the sign, or chirality, of the DMI can be changed by the helicity of the laser pulse.

\section{Detecting the chirality of the signal in the output arm}

\begin{figure}[!h]
	\centering\includegraphics[scale=0.5, angle=270]{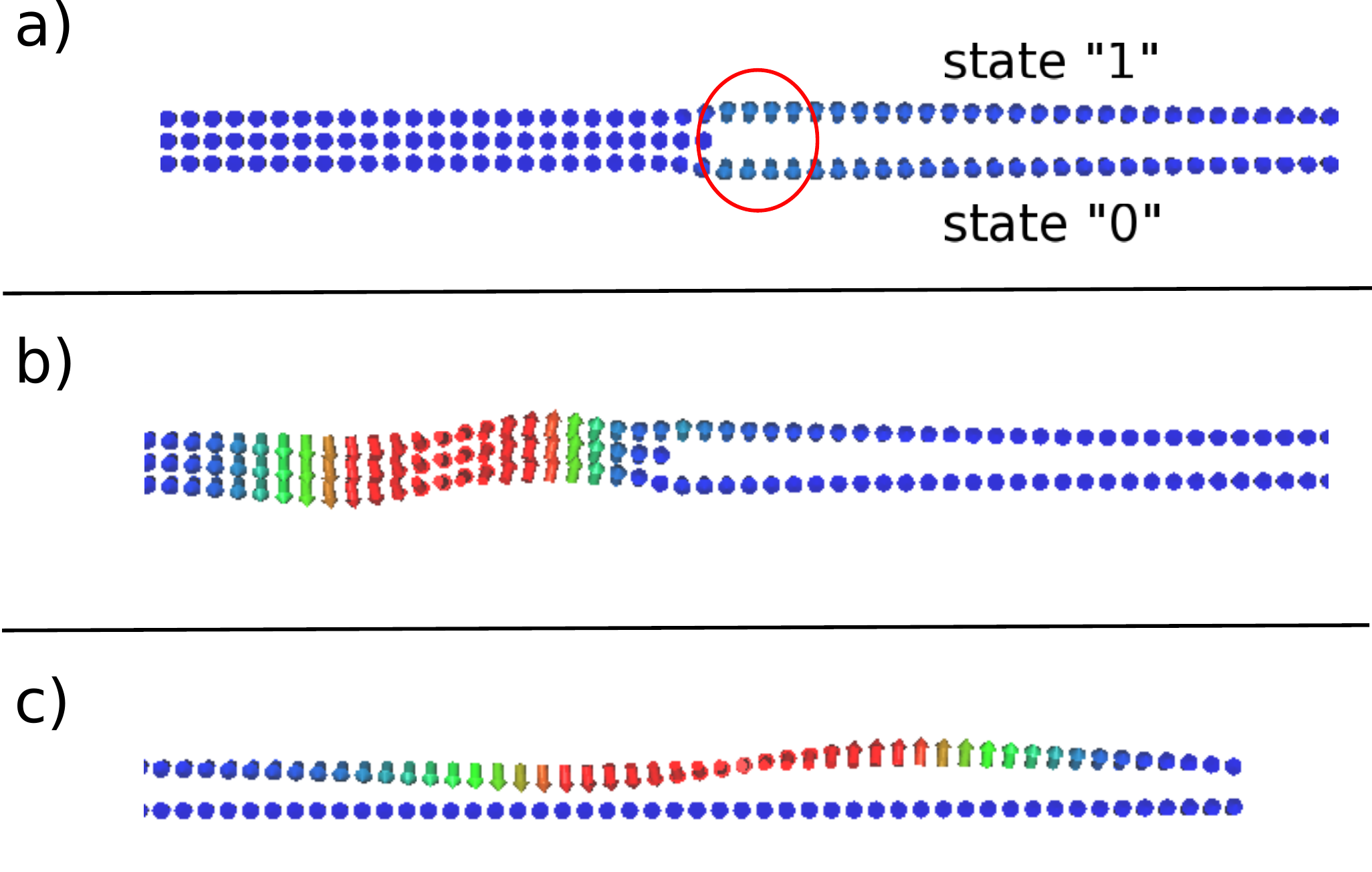}
	\caption{\label{double_output} Device with the double output arm geometry \textbf{a)}, that transmits a logical state "1" in the upper arm and logical state  "0" in the lower arm. 
		 \textbf{b)} Simulations of the resulting output soliton with state "1" in the interacting region, just before entering the two output lines. \textbf{c)} Snapshot of the  "1"  state soliton, which is found to travel only in the upper arm.}
\end{figure}

In the main text of the paper we eluded to different ways to detect the chirality of the signal of the output arm. We provide here a very reliable way in which to do this.
The only difference compared to the device geometry shown in Fig.1 of the main paper, is that we now consider an interacting region that connects the three input arms and two output arms. The two output arms have different signs of the DM vector (shown in Fig.~\ref{double_output}a), although the strength of the DM vector is the same as for the input arms, i.e. $D=0.2$ mRy. Since the sign of the DMI strength is different in the two output arms one might expect that only solitons with a certain chirality are allowed to travel in each arm, something our simulations support. We performed a large number of simulations of the three input majority device, in the geometry shown in  Fig.~\ref{double_output}a. In these simulations the strength of the Heisenberg exchange interaction in the two output arms was choosen to be the same as for the input arms, i.e. $J=1$ mRy. With extremely high reliability we found that solitons which are in state "1" survive only in the upper arm (e.g. as shown in Fig.~\ref{double_output}c and Supplementary Movie 3) and the  "0" state solitons can survive only in the lower arm (Supplementary Movie 4). 


\noindent
This device shown in Fig.~\ref{double_output}a makes the detection of the chirality of the output signal very easy, since one knowns that a soliton detected in the upper arm must have the output state is "1", and a logical state "0" if the signal comes from the lower arm.

\section{Numerical aspects of the functionality of the proposed device}

In our simulations we use realistic values of the magnetic interactions of both output and  input arms, more specifically we have worked with exchange coupling $J=1$ mRy and DMI strength $D=0.2$ mRy. However, results of our numerical simulations reveal that the device in Fig. 1, of the main text, is somewhat sensitive to the chosen parameters of the system: for example, when $J^\prime$ exceeds a certain critical value of about 0.3 mRy the device looses its functionality, and the majority gate does not give an output signal. In order to illustrate this sensitivity, we provide in Table \ref{thresholds} quantitative values of the threshold current density for three different values of $J^\prime$, needed to be applied to detect a soliton in the output arm. 
Note that, the results on the threshold current densities are very close to each other, meaning that there is almost no dependence on $J^\prime$ of the output arm on receiving the output signal. It is obvious though, that these values are much higher comparing to the threshold current density which makes solitons move and is approximately $j= 2.8 \times 10^8$ A/m$^2$. The latter was intuitively expected and associated with a need of flipping magnetic moments in the absence of DMI. 

\begin{table}[h]
	\begin{center}
		\caption{\label{thresholds} Threshold currents for readable output with respect to the Heisenberg exchange interaction of the output branch $J^\prime$.}
		\begin{tabular}{|c|c|}
			\hline \hline
			Output exchange interaction $J^\prime$ & Threshold current density \\
			\hline \hline
			0.1 mRy & 5 $\times$ 10$^{12}$ A/m$^2$ \\
			0.2 mRy & 4.8 $\times$ 10$^{12}$ A/m$^2$ \\
			0.3 mRy & 6.7 $\times$ 10$^{12}$ A/m$^2$ \\
			\hline
			\hline
		\end{tabular}
	\end{center}
	
\end{table}